\documentstyle[12pt]{article}
\advance \topmargin by -\headheight
\advance \topmargin by -\headsep
     
\evensidemargin \oddsidemargin
\def\rf#1{(\ref{eq:#1})}
\def\lab#1{\label{eq:#1}}
\def\nonu{\nonumber}
\def\br{\begin{eqnarray}}
\def\er{\end{eqnarray}}
\def\be{\begin{equation}}
\def\ee{\end{equation}}

\def\lb{\lbrack}
\def\rb{\rbrack}

\def\({\left(}
\def\){\right)}
\def\v{\vert}                     

\def\bc{\begin{center}}
\def\ec{\end{center}}

\newcommand{\sect}[1]{\setcounter{equation}{0}\section{#1}}

\relax
\def\tr{\mathop{\rm tr}}                  
\def\Tr{\mathop{\rm Tr}}                  

\newcommand\sbr[2]{\left\lbrack\,{#1}\, ,\,{#2}\,\right\rbrack} 
\def\a{\alpha}
\def\b{\beta}
\def\c{\chi}
\def\d{\delta}

\def\eps{\epsilon}

\def\h{{1\over 2}}
\def\l{\lambda}

\def\o{\over}

\def\p{\phi}
\def\P{\Phi}
\def\pa{\partial}

\def\pr{\prime}

\def\s{\sigma}

\def\Th{\Theta}

\def\cA{{\cal A}}
\def\cB{{\cal B}}
\def\cC{{\cal C}}

\def\cG{{\cal G}}
\def\cH{{\cal H}}

\def\cJ{{\cal J}}
\def\cK{{\cal K}}
\def\cL{{\cal L}}

\def\cW{{\cal W}}

\def\lie{{\cal G}}

\font \msb=msbm10 scaled \magstep1

\newcommand{\IC}{\mbox{\msb C} }

\newcommand{\IZ}{\mbox{\msb Z} }

\newtheorem{definition}{Definition}[section]

\def\proof{\par{\it Proof}. \ignorespaces} \def\endproof{{$\Box$}\par}

\def\Win1{{\bf W_{1+\infty}}}           

\def\win1{{\bf w_{1+\infty}}}

\newcommand\fourmat[4]{\left(\begin{array}{cc}  
{#1} & {#2} \\ {#3} & {#4} \end{array} \right)}

\def\vp{{\varphi}}

\newcommand{\ct}[1]{\cite{#1}}
\newcommand{\bi}[1]{\bibitem{#1}}
\def\ba{\be\begin{array}{c}}
\def\ea{\end{array}\ee}
\def\kere{\mbox{\rm Ker (ad $E$)}}

\def\cgh{{\widehat {\cal G}}}

\begin{document}

\begin{titlepage}
\vspace{-1cm}
\noindent
\vskip .3in

\begin{center}
{\large\bf Integrable Hierarchy for }
\end{center}
\begin{center}
{\large\bf Multidimensional Toda Equations and    }
\end{center}
\begin{center}
{\large\bf Topological-Anti-topological Fusion}
\end{center}
\normalsize
\vskip .4in

\begin{center}
 H. Aratyn

\par \vskip .1in \noindent
Department of Physics \\
University of Illinois at Chicago\\
845 W. Taylor St.\\
Chicago, Illinois 60607-7059\\
\par \vskip .3in

\end{center}

\begin{center}
J.F. Gomes
and A.H. Zimerman

\par \vskip .1in \noindent
Instituto de F\'{\i}sica Te\'{o}rica-UNESP\\
Rua Pamplona 145\\
01405-900 S\~{a}o Paulo, Brazil
\par \vskip .3in

\end{center}

\begin{center}
{\large {\bf ABSTRACT}}\\
\end{center}
\par \vskip .3in \noindent

The negative symmetry flows are incorporated into the Riemann-Hilbert
problem for the homogeneous $A_m$-hierarchy and its
$\widehat{gl} (m+1, \IC)$ extension.

A loop group automorphism of order two is used to define 
a  sub-hierarchy of $\widehat{gl} (m+1, \IC)$ hierarchy
containing only the odd symmetry flows.
The positive and negative flows of the $\pm 1$ grade coincide with 
equations of the multidimensional Toda model
and of topological-anti-topological fusion.

\end{titlepage}

\sect{Introduction}

Integrable models and their multi-time formulation find a natural and 
universal setting in terms of the Riemann-Hilbert problem.  
According to the Riemann-Hilbert problem 
a loop group element $G(\l )$ factorizes as
\be
G(\l ) =  G_{-} (\l ) G_{+}(\l )
\lab{ggg}
\ee
where $G_{+} (\l ), \;  (G_{-}(\l ))$ belong to subgroups
constructed from  positive (strictly negative)
$\cgh_{+}, \; (\cgh_{-})$ graded subalgebras, respectively.
The gradation is defined by powers of the spectral parameter $\lambda$
counted by the grading operator $d = \lambda {{d }\o
{d\lambda}}$. Such a gradation is known as homogeneous
gradation.
The loop algebra  $\cgh$ decomposes  into graded
subspaces $ \cgh= \oplus_{n \in \IZ} \, \cgh_n$ with  $\cgh_n$
such that $\lb d \, , \, \cgh_n \rb = n\, \cgh_n$.
The parameter $\lambda$ plays a two-fold role; it appears
as a spectral parameter in the fundamental linear spectral problem of the integrable model
and also serves as a loop variable parametrizing the closed contour on the 
complex plane taken here to be a unit circle $S^1$.

The matrices appearing in the Birkhoff factorization \rf{ggg} 
are linked to another important concept in the
soliton theory, namely the dressing transformation which maps a vacuum
to soliton solutions.
In the context of the generalized Drinfeld-Sokolov formalism the dressing
transformation introduces a multi-time structure associated
to the positive grade  Heisenberg subalgebra in $\cgh$
\ct{Drinfeld:1984qv,DeGroot:1992ca,Burroughs:1991bd}.
As we vary the positive grade, the isospectral times corresponding to the generators
of the Heisenberg subalgebra form an hierarchy of equations of motion.
This provides a standard algebraic derivation of the integrable hierarchy
which can be embedded within the Riemann-Hilbert problem.
The combination of these two basic concepts allows us to derive
all fundamental objects like the Hamiltonian densities and the tau function
using the flows and algebraic structure inherited from the Riemann-Hilbert
problem.

The well-known $\lie=sl(2) $ examples of the dressing method are  mKdV  and AKNS
hierarchies associated to the principal and homogeneous gradations, respectively.
Also, corresponding to $\lie=sl(2) $ with the principal and homogeneous 
gradations are the sine-Gordon and complex sine-Gordon hierarchies.
However they fall outside the scope of
the standard dressing technique because their times are associated
to the negative grade generators of the Heisenberg subalgebra.
This motivates construction of a formalism
which would incorporate both positive and negative times.
The Riemann-Hilbert problem naturally allows 
for such a generalization \ct{dorfmeister,Aratyn:2000wr}.
The outcome of this construction is a unified framework
with hierarchies of evolution equations corresponding to mutually
commuting positive and negative flows.

In this paper, we generalize the multi-time formulation
of the Riemann-Hilbert problem by including all underlying
self-commuting symmetry flows of the system for the 
$\widehat{gl} (m+1)$ loop algebras.
The isospectral times are flows corresponding  to  the Heisenberg 
subalgebra generators $E^{(k)} =
\mu_m \cdot H^{(k)} $ with positive and negative grade ($k \in \IZ$), where
$\mu_m$ is the $m$-th fundamental weight.
It is known \ct{Aratyn:1997ji}, that in the homogeneous gradation the 
centralizer of the Heisenberg subalgebra,
$ \hat{gl} (m) \times \hat{u} (1)$, is non-abelian for $m>1$.
The symmetry structure in question is given by
the self-commuting symmetry flows associated to the abelian 
generators $ E^{(k)}_{jj}=\lambda^k E_{jj}, j=1,{\ldots} ,m+1$
of the centralizer of the Heisenberg subalgebra.
Here, we use a notation $ (E_{rs})_{ij} = \delta_{ir} \delta_{js}$.
The presence of both positive and negative
sectors of the extended hierarchy agrees with a complex structure
of $\widehat{gl} (m+1, \IC)$ symmetry with flows 
generated by $ E^{(k)}_{jj}, j=1,{\ldots} ,m+1$.
When we consider the positive flows only,  
the above structure reduces naturally 
to the homogeneous $A_m=\widehat{sl} (m+1)$-hierarchy which 
generalizes the AKNS hierarchy
for $m>1$.

The diagonal generators of  $\widehat{gl} (m+1)$ generate the multi-dimensions
of the Toda model as flows with $\pm 1$ gradations of the underlying 
hierarchy.
Those flows take a form of the Cecotti-Vafa equations of the 
topological-anti-topological fusion \ct{Cecotti:1991me,Dubrovin:1993yd}
when considered within the sub-hierarchy restricted (or twisted) 
by a specific loop group automorphism.
This unveils the topological field theory concepts in the context of
the reduction of the homogeneous
$\widehat{gl} (m+1, \IC)$-hierarchy. 
Similar integrable structure with positive flows only was recently used to 
find solutions to the WDVV equations \ct{Aratyn:2001cj}.

In Section 2, we define a Riemann-Hilbert problem for
the integrable model with the underlying $\widehat{gl} (m+1)$
loop algebra
with the homogeneous gradation
containing positive and negative $\widehat{gl} (m+1)$
symmetry flows.
For the model with only positive multi-times this hierarchy
reduces to the homogeneous $A_m$-hierarchy.
In Section 3, we study the action of associated commuting symmetry flows on the
dressing matrices of
positive and negative gradation and derive the conservation laws and
expressions for the Hamiltonian densities.
The underlying tau function is given by taking an expectation value of the 
Riemann-Hilbert equation based on the highest weight vacuum of the associated Kac-Moody algebra
\ct{Wilson:1993}.
We also discuss the positive dressing matrix $M$ and its
inverse using the relation between the algebraic and pseudo-differential 
approaches.

In Section 4, we derive the multidimensional Toda model
equations from the positive and negative flows of $\pm 1$ grade 
of the $\widehat{gl} (m+1, \IC)$-hierarchy.
Next, we impose the set of constraints on the dressing matrices 
defining a consistent sub-hierarchy of $\widehat{gl} (m+1,
\IC)$-hierarchy allowing only odd positive and negative flows.
The dressing matrices are constrained to be the fixed points
of a specific loop group automorphism of order two.
The Cecotti-Vafa equations of topological-anti-topological fusion
are found among the positive and negative flows of $\pm 1$ grade of the
reduced integrable sub-hierarchy.
The similar sub-hierarchy (without the negative flows)
has recently been shown 
to provide solutions of the Darboux-Egoroff system of 
PDE's \ct{Aratyn:2001cj}.
Here, due to the presence of negative and positive flows, we 
obtain two coupled Darboux-Egoroff systems embedded in the complex-like
structure of the Cecotti-Vafa equations.
As an example we discuss the extended AKNS/complex sine Gordon model
and its reduction.

\sect{Extended Riemann-Hilbert Problem and ${\bf \widehat{gl} (m+1)}$
Symmetry Flows}  
  
We will introduce the Riemann-Hilbert problem in terms two subgroups of 
the Lie loop group $G$ defined as :
\br
G_{-} &=& \left\{ g \in G \v g(\l) = 1 + \sum_{i<0} g^{(i)}  \right\}
\lab{g-neg}\\
G_{+} &=& \left\{ g \in G \v g(\l) = \sum_{i\geq0} g^{(i)}  \right\}
\lab{g-pos}
\er
where $g_i$ has grading $i$ with respect to a homogeneous gradation
defined by derivation $d = \l d /d \l$.
It also holds that $G_{+} \cap G_{-} = I$.
Let the loop algebra corresponding  to $G$ be $\cgh = 
{\widehat {gl}} (m+1)$.
This algebra splits into the direct sum $\cgh=\cgh_{+} \oplus\cgh_{-}$
with respect to the homogeneous gradation, where $\cgh_{\pm}$ are Lie
algebras associated with the subgroups $G_{\pm}$.

We now define an extended Riemann-Hilbert factorization problem for the
homogeneous gradation :
\begin{equation}
\exp \({\sum_{j=1}^{m+1} \sum_{n=1}^{\infty} E^{(n)}_{jj}u^{(n)}_j}\) \,  
g \,
\exp \({-\sum_{j=1}^{m+1} \sum_{n=1}^{\infty} E^{(-n)}_{jj}u^{(-n)}_j} \)
= \Theta^{-1} ({\bf u},\l) \,  M  ({\bf u},\l) 
\lab{rh-def}
\end{equation}
where $g$ is a constant element in $G_{-}G_{+}$
while $ \Theta^{-1} \in G_{-}$, $M \in G_{+}$.
We use the multi-time notation with
$({\bf u})= ({\bf u_1}, {\ldots} , {\bf u_{m+1}})$
to denote $m+1$ multi-flows ${\bf u_j}$. Each argument
${\bf u_j}$, $j=1,{\ldots} ,m+1$  is an abbreviated notation for the 
multi-flows $u_j^{(n)}$ with $n$ between $-\infty$ to $\infty$.

The exponential term $\exp \({-\sum_{j=1}^{m+1}
\sum_{n=1}^{\infty} E^{(-n)}_{jj}u^{(-n)}_j}\)$ on the left hand side of 
\rf{rh-def} contains terms of negative grade. This term 
extends the standard Riemann-Hilbert problem of the KP like models
by negative flows.
Acting with $\frac{\partial}{\partial u^{(n)}_j}$
on both sides of \rf{rh-def} we find
\begin{equation} 
 \Theta (u)  E^{(n)}_{jj} \Theta^{-1} (u) = \Theta(u) 
\left( \frac{\partial}{\partial u^{(n)}_j} \Theta^{-1} (u) \right)
+\left( \frac{\partial}{\partial u^{(n)}_j} M (u) \right)
 M^{-1}(u) \, .
\lab{rh-tka}
\end{equation}
Note, that $\Theta
\left( \frac{\partial}{\partial u^{(n)}_j} \Theta^{-1} \right)$ is in 
${\cal G}_{-}$
and $\left( \frac{\partial}{\partial u^{(n)}_j} M  \right)
 M^{-1}$ is in ${\cal G}_{+}$.
Hence, for the positive symmetry flows one derives from \rf{rh-tka}
expressions:
\br
\frac{\partial}{\partial u^{(n)}_j} \Theta ({\bf u}, \lambda) 
&=& - \left(\Theta E^{(n)}_{jj} \Theta^{-1} 
\right)_{-} \Theta ({\bf u}, \lambda) 
\lab{uthpos}\\
\frac{\partial}{\partial u^{(n)}_j} M ({\bf u}, \lambda) 
&=&  \left(\Theta E^{(n)}_{jj} \Theta^{-1} 
\right)_{+} M ({\bf u}, \lambda) 
\lab{ummpos}
\er
where $({\ldots} )_{\pm}$ denote the projections into $\cgh_{\pm}$.
{}From equation \rf{uthpos} we find the tracelessness condition
\be
\sum_{j=1}^{m+1} \frac{\partial}{\partial u^{(n)}_j} \Theta ({\bf u}, \lambda) 
=0 
\lab{traceum}
\ee
which allows to consider the positive flows together with 
the negative dressing matrix
$ \Theta ({\bf u}, \lambda) $ as a homogeneous $A_m$-hierarchy
with a $\widehat{sl} (m+1)$ symmetry of flows. For $m=1$ we recover in this 
way the AKNS hierarchy.

After applying $\frac{\partial}{\partial u^{(-n)}_j}$ on both sides 
of \rf{rh-def} one finds for the negative flows :
\br
\frac{\partial}{\partial u^{(-n)}_j} \Theta ({\bf u}, \lambda) 
&=&  \left( M E^{(-n)}_{jj} M^{-1} 
\right)_{-} \Theta ({\bf u}, \lambda) 
\lab{uthneg}\\
\frac{\partial}{\partial u^{(-n)}_j} M ({\bf u}, \lambda) 
&=& - \left( M E^{(-n)}_{jj} M^{-1} 
\right)_{+} M ({\bf u}, \lambda) 
\lab{ummmin}
\er
One observes that the above system of flow equations is invariant under
the right multiplication of $M ({\bf u}, \lambda) $
by a constant diagonal matrix $H$.

Similarly, we find the tracelessnes condition :
\be
\sum_{j=1}^{m+1} \frac{\partial}{\partial u^{(-n)}_j} M ({\bf u}, \lambda) 
=0 
\lab{tracedois}
\ee
which allows to consider the negative flows together with the positive dressing matrix
$ M ({\bf u}, \lambda) $ as a homogeneous $A_m$-hierarchy,
which generalizes the complex sine-Gordon equation hierarchy \cite{Aratyn:2000wr}.

For the hierarchy containing both positive and negative flows
and the dressing matrices $ \Theta ({\bf u}, \lambda) $ 
and  $M ({\bf u}, \lambda) $ 
we work with the full $\widehat{gl} (m+1)$  symmetry.

We now adress the issue of commutativity of the flows.
Applying, respectively $\pa^2 / \pa u^{(\pm n)}_j \pa u^{(\pm k)}_i$ 
and $\pa^2 / \pa u^{(\pm k)}_i \pa u^{(\pm n)}_j $ on both
sides of eq.\rf{rh-def} produces identical results due to commutativity
of $E^{(\pm n)}_{jj}$ with $E^{(\pm k)}_{ii}$. 
This ensures that :
\be
{\pa^2 \Theta ({\bf u}) \o \pa u^{(\pm n)}_j \pa u^{(\pm k)}_i} 
= {\pa^2 \Theta ({\bf u}) \o \pa u^{(\pm k)}_i \pa u^{(\pm n)}_j } 
\;\; ; \; \; {\pa^2 M ({\bf u})\o \pa u^{(\pm n)}_j \pa u^{(\pm k)}_i} 
= {\pa^2 M ({\bf u}) \o \pa u^{(\pm k)}_i \pa u^{(\pm n)}_j } 
\lab{tntk}
\ee
In a mixed case we apply $\pa^2 / \pa u^{(n)}_j \pa u^{(- k)}_i$ 
and $\pa^2 / \pa u^{(- k)}_i \pa u^{( n)}_j $ on both
sides of eq.\rf{rh-def}. As before, both operations give identical
results although this time the outcome follows automatically from 
the construction and does not rely on the commutativity of 
$E^{(n)}_{jj}$ with $E^{(- k)}_{ii}$. 
Hence it holds that :
\be
{\pa^2 \Theta ({\bf u}) \o \pa u^{(n)}_j \pa u^{(- k)}_i} 
= {\pa^2 \Theta ({\bf u}) \o \pa u^{(- k)}_i \pa u^{( n)}_j } 
\;\; ; \; \; {\pa^2 M ({\bf u}) \o \pa u^{(n)}_j \pa u^{(- k)}_i} 
= {\pa^2 M ({\bf u}) \o \pa u^{(- k)}_i \pa u^{( n)}_j } 
\lab{tntmk}
\ee

The isospectral flows defining conservation laws of the underlying
integrable system are related to
$u_{m+1}^{(k)}$-flows :
\be
{\pa \o \pa t_k} =  {\pa \o \pa  u_{m+1}^{(k)}  }
\lab{iso-def}
\ee
with identification being valid for both positive and negative $k$.
  
The standard dressing construction \ct{Aratyn:1997ji} associates 
isospectral flows to a semisimple grade-one element $E^{(1)}$ :
\be
E^{(1)} = \l E= \mu_m \cdot H^{(1)}=\frac{\lambda  }{ m+1} I - \lambda E_{m+1 \, m+1}
\lab{defE}
\ee
where $\mu_m$ is the m-th fundamental weight of $sl (m+1)$.
The kernel of ${\rm ad}\, E$ is :
\be
\cK= \kere = \{ {\widehat {gl}} (m) \oplus {\hat u}(1) \}
\ee
with ${\hat u}(1)$ generating the center $\cC (\cK)
= \{ \mu_m \cdot H^{(k)}\,\, , \,\,   k \in \IZ \}$ of $\kere$.
Each isospectral flow $t_k$ is assigned to an element :
\be
E^{(k)} \equiv \mu_m \cdot H^{(k)} \, , \, \qquad \mbox{\rm $k \geq 1$ }
\ee
in the center $\cC (\cK)$ according to:
\be
{\pa \o \pa t_k} \Th = (\Th  E^{(k)} {\Th}^{-1})_{-}\, \Th 
\lab{Th-flowsa}
\ee
which coincides with \rf{uthpos} for $j =m+1$ in agreement with the 
definiton \rf{iso-def}.

For $k=1$, we obtain from \rf{uthpos}
\begin{equation}
\left(\frac{\partial}{\partial u^{(1)}_{m+1}} - E^{(1)}_{m+1\,m+1} -
\lbrack \theta^{(-1)}, E^{(1)}_{m+1\,m+1} \rbrack \right) \Theta = 
-\Theta E^{(1)}_{m+1\,m+1} \, , 
\lab{thempo}
\end{equation}
where $\theta^{(-1)}$ is a term of expansion of
$\Theta = 1 + \theta^{(-1)} + O(\l^{-2})$.
Equation \rf{thempo} can also be rewritten as :
\begin{equation}
\left(\frac{\partial}{\partial t_1} + E^{(1)} +
\lbrack \theta^{(-1)}, E^{(1)} \rbrack \right) \Theta = \Theta E^{(1)} \, , 
\lab{theone}
\end{equation}
in agreement with \rf{Th-flowsa}.

The potential $A\equiv \lbrack \theta^{(-1)}, E^{(1)} \rbrack $ lies in the 
grade-zero component of the image of $ad (E)$ and can therefore be parametrized 
as :
\begin{equation}
A  = \lbrack \theta^{(-1)}\, , \,E^{(1)}  \rbrack 
\,= \,-\lbrack \theta^{(-1)}\, , \,E^{(1)}_{m+1\,m+1}   \rbrack 
\,= \,\sum_{i=1}^{m} \left( - \Psi_i \, 
E_{i \, m+1} + \Phi_i \, E_{m+1 \, i}\right)
\, .
\lab{a-homo}
\end{equation}
We will refer to the hierarchy defined by the isospectral times
from \rf{Th-flowsa} and symmetry flows from equation \rf{uthpos}
with the parametrization in \rf{a-homo}
as the homogeneous $A_m$-hierarchy \ct{Fordy:1983ax,Flaschka:1983ac,Aratyn:1995bm}.

Given the parametrization \rf{a-homo}, equation \rf{thempo} can be cast
into the form 
\be
\Theta^{-1}\; L \; \Theta  = \pa_x - E^{(1)}_{m+1\,m+1}
\lab{thlth} 
\ee
involving the matrix Lax operator $L$ :
\be
L = \pa_x - E^{(1)}_{m+1\,m+1} + A
\lab{matlaxdef}
\ee
where $\pa_x = \pa / \pa t_1 = \pa / \pa u^{(1)}_{m+1}$ is 
acting to the right as an operator according to the Leibniz rule.
In \ct{Aratyn:2000sm}, the homogeneous $A_m$ hierarchy
was constructed by rotating matrices of the Lax operator 
\rf{matlaxdef}
into the kernel of $ad (E^{(1)})$.

The linear spectral problem emerges when 
setting $n=1$ and $j=m+1$ in \rf{ummpos}. This reveals the 
matrix $M$ as a solution of \ct{Aratyn:2000wr}:
\be 
L (M) =\( \pa_x - E^{(1)}_{m+1\,m+1} + A\) (M) =0
\lab{lmo}
\ee

\sect{The Homogeneous $A_m$-Hierarchy from the Riemann-Hilbert
Factorization}  

\subsection{Isospectral Flows, Hamiltonians}

We will now discuss the conservation laws of the
homogeneous $A_m$-hierarchy. These laws are based on requirement of 
locality with respect to the potentials $\Phi_i, \Psi_i$ from \rf{a-homo}.
A quantity is considered local if it is a polynomial of $\Phi_i, \Psi_i$ 
and their derivatives.

{}From eq. \rf{uthpos} we find:
\be
\frac{\partial}{\partial u^{(n)}_i} (\Th  E^{(n^{\pr})}_{jj} {\Th}^{-1})
= \lb \, \Th  E^{(n^{\pr})}_{jj}{\Th}^{-1}\, , \, 
(\Th  E^{(n)}_{ii} {\Th}^{-1})_{-}\, \rb 
\lab{Th-flowsc}
\ee
and therefore:
\br
&&\frac{\partial}{\partial u^{(n)}_i} 
{\rm Tr}_0 \( E^{(1)} {\Th} E^{(n^{\pr})}_{jj} {\Th}^{-1} \) - 
\frac{\partial}{\partial u^{(n^{\pr})}_j}  {\rm Tr}_0 
\( E^{(1)} {\Th} E^{(n)}_{ii} {\Th}^{-1} \)
\nonu \\
&=& {\rm Tr}_0 \( E^{(1)} \lb \, \Th  E^{(n^{\pr})}_{jj}  {\Th}^{-1}\, , \, 
\Th  E^{(n)}_{ii} {\Th}^{-1}\, \rb \) =0
\lab{hnhm}
\er
with the trace which includes projection on the $\l^{0}$ term:
\be
{\rm Tr}_0 (XY) \equiv \langle X, Y \rangle_0 = \sum_{i+j=0} \tr (X_iY_j) \;\;,
\quad X= \sum_i X_i \l^i \; , \;\; Y= \sum_i Y_i \l^i
\lab{tr-def}
\ee
It is therefore natural to associate to the Riemann-Hilbert factorization approach
the quantities:
\be 
 \cH^{(n)}_j = \,   {\rm Tr}_0 \( E^{(1)} {\Th} E^{(n)}_{jj} {\Th}^{-1} \) 
 ,\quad j=1,{\ldots} , m+1
 \lab{hamsi} 
\ee
which satisfy according to \rf{hnhm} 
the identity 
\be
{\pa \cH^{(n)}_j \o \pa u^{(n^{\pr})}_k} - {\pa \cH^{(n^{\pr})}_k\o \pa u^{(n)}_j } 
=0, \quad n, n^{\pr} >0
\lab{hn-id}
\ee
Furthermore, $ \cH^{(n)}_j $ appears to be a dervative $  \cH^{(n)}_j = 
\pa_x \cJ^{(n)}_j $ of a current density :
\be
\cJ^{(n)}_j = -
{\rm Tr}_0 \( \l {d  \Theta \o d \l}    E^{(n)}_{jj} {\Theta}^{-1}  
  \) = -
{\rm Res}_{\l} \(\tr \(  E^{(n)}_{jj} {\Theta}^{-1}  
{d  \Theta \o d \l} \)  \) , \; \, j=1, {\ldots} ,m+1
\lab{jnj-tau}
\ee 
which satisfies an identity :
\be
{\pa \cJ^{(n)}_j \o \pa u^{(n^{\pr})}_k} - {\pa \cJ^{(n^{\pr})}_k\o \pa u^{(n)}_j } 
=0, \quad n, n^{\pr} >0
\lab{hn-id-gen}
\ee
analogous to \rf{hn-id}.

The quantity 
\be
{\pa \cJ^{(n)}_j \o \pa u^{(n^{\pr})}_k}  =
{\rm Tr}_0 \( \l {d  (\Th E^{(n^{\pr})}_{kk}\Th^{-1})_{+} \o d \l}\, 
\Theta \, E^{(n)}_{jj}\, {\Theta}^{-1} \) 
\lab{panjm}
\ee
becomes local (in terms of $\Psi_i, \Phi_i$) for $j=k=m+1$, as observed in
\ct{Aratyn:2000sm}.
We therefore obtain the local conservation laws in terms of
\be
\cH_n = \cH^{(n)}_{m+1} =
-  {\rm Tr}_0 \( E^{(1)} {\Th} E^{(n)} {\Th}^{-1} \) 
 \lab{hams} 
\ee
and
\be
 \cJ_n=  \cJ^{(n)}_{m+1}  =
 {\rm Tr}_0 \( \l {d  \Theta \o d \l} E^{(n)} {\Theta}^{-1} \) 
= {\rm Res}_{\l} \(\tr \(  E^{(n)} {\Theta}^{-1}  {d  \Theta \o d \l} \)  \)
\lab{jns} 
\ee
connected via $\cH_n = \pa_x  \cJ_n$.
The observation that \rf{panjm} becomes local for $j=k=m+1$
provides a direct way to prove conservation of the 
Hamiltonians $ H_n = \int \cH_n \; dx $. One 
notices that
\be 
{\pa \o \pa t_{n^{\pr}}} H_n = \int \pa_x {\pa \o \pa t_{n^{\pr}}} \cJ_n = 0
, \quad n, n^{\pr} >0
\lab{dtmhn}
\ee
due to locality of the relevant expression in \rf{panjm}
for $j=k=m+1$.

Consider the following expansions :
\br
\Theta &= &1 + \sum_{k=1}^{\infty} \theta^{(-k)} / \l^k
\lab{thetexp}\\
\Theta E^{(1)}  {\Theta}^{-1} &=& 
E^{(1)} +A +\sum_{k=1}^{\infty} A^{(-k)} / \l^k
\lab{theth}\\
\Theta E^{(n)} {\Theta}^{-1} &=& \l^{n} E + \l^{n-1} A +\sum_{k=1}^{\infty}
\l^{n-k-1} A^{(-k)} \lab{thenth}
\er
Plugging $j=k=m+1$ and $n^{\pr}=1$ into the identity \rf{hn-id-gen} 
yields :
\be
\tr \( E \,A^{(-n)} \) = - \h \sum_{k=0}^{n-1} \tr \(  A^{(-k)} A^{(1+k-n)} \)
\;,\; \; n \geq 1
\lab{recurb}
\ee
In terms of the elements of expansions \rf{thetexp}-\rf{thenth} 
$ \cH_n $ takes the form 
\be
\cH_n =  - \tr \( E A^{(-n)}\)  = 
- \h \sum_{k=0}^{n-1} \tr \(  A^{(-k)} A^{(1+k-n)} \)
\;  \; n \geq 1
\lab{recurbb}
\ee
which allow us to cast Hamiltonians into the well-known  form 
due to \ct{Flaschka:1983ac} :
\be
\cH_n =  \h {\rm Tr}_0 (\l^{n+1} X^2 (\l) )  \;\; \;{\rm for}\,\;
n\geq 1
\lab{fla}
\ee
where
\be
X (\l) = \sum_{i=1}^{\infty} X_i \l^{-i} = \sum_{k=1}^{\infty}
A^{(-k+1)} \l^{-k}= \( \Theta E^{(0)}  {\Theta}^{-1}\)_{-}
\lab{x-exp}
\ee
where $A^{(0)}$ denotes the potential $A$. 
$\cH_n$ can also be written in an equivalent form :
\be
\cH_n =  \h \langle \l^n X, X \rangle_{-1} = \h \oint d \l \l^n \tr (X^2)
\lab{ham-mone}
\ee
where we introduced another symmetric bilinear form:
\be
\langle X, Y \rangle_{-1} \equiv  {\rm Res}_{\l}
\( \tr (XY) \) = \sum_{i+j=-1} \tr (X_iY_j) 
\lab{bil-form}
\ee
{}From:
\be
{d \o d \eps} \h \langle \l^n (X+\eps Y)^2 \rangle_{-1} \v_{\eps=0} 
= \langle \l^n X, Y \rangle_{-1} = \langle \nabla \cH_n, Y \rangle_{-1}
\lab{dhn}
\ee
we identify $\nabla \cH_n = \l^n X$  
which according to the Adler-Kostant-Symes formalism yields for
the flows:
\be
{\pa  X (\l) \o \pa  t_n}  =  \lb \( \l^{n} X(\l)\)_{-} ,  X (\l) \rb
= \lb \({\Th} E^n {\Th}^{-1} \)_{-} , X (\l) \rb
\lab{fla-flows}
\ee
or equivalently
\be
{\pa X (\l) \o \pa t_n}  =  - \lb \( \l^{n} X(\l)\)_{+} ,  X (\l) \rb_{-}
\lab{fl-flows}
\ee
which fully agrees with the dressing formula in \rf{Th-flowsc}.

\subsection{The Tau Function}

The identity \rf{hn-id-gen} suggests that $ \cJ^{(n)}_j $
can be written as $\pa / \pa u^{(n)}_j$ of
some function of the $\Phi_i, \Psi_i$ variables.
Conventionally, this function is denoted as the logarithm of the $\tau$-function.
Accordingly, the $\tau$-function is introduced 
by the relation :
\be
\cJ^{(n)}_j = - {\pa  \log \tau  \o \pa u^{(n)}_j}
\lab{jnjta}
\ee
with more conventional identity 
\be    
\cJ_n= 
\, - {\pa \o \pa  t_n} \log \tau  \,.
\lab{jn-tau}
\ee
being a special case corresponding to $j=m+1$.
Note, that the relation
\be
 {\pa  \cJ_n \o \pa t_ {n^{\pr}}} - {\pa  \cJ_ {n^{\pr}} \o \pa t_n} \,=\, 0 \,
, \quad n,  {n^{\pr}} > 0
\lab{jn-id}
\ee
also holds as a special case of \rf{hn-id-gen}.

We will use below the setting of $A_m$ Kac-Moody algebra 
to integrate these equations to obtain a closed
expression for the $\tau$-function. 

Let the elements of the Kac-Moody algebra be $\xi + s {\hat k} $
where $ \xi (\varphi) \, \equiv \,  \xi^a (\varphi) T_a $ is a function on $S^1$ 
with values in the underlying finite-dimensional Lie algebra $\lie=sl (m+1)$ 
and $s$ is the central element.
The Kac-Moody algebra reads :
$$
\lbrack \xi_1 + s_1 {\hat k} \, , \, \xi_2 + s_2 {\hat k} \rbrack =
\lbrack \xi_1 \, , \, \xi_2 \rbrack\, + \, {\hat k} \omega \(\xi_1 
\, , \, \xi_2\) =
\lbrack \xi_1 \, , \, \xi_2 \rbrack
\, + \, {\hat k} \oint {d \varphi \o 2 \pi  } \tr \( \pa_\varphi \xi_1 \xi_2 \) 
$$
where $\varphi$ is an $S^1$ angle variable.
The adjoint action of the group $G$ on $\lie$ 
$$
Ad(g) \, \xi = g \xi g^{-1}
$$
generalizes to
\be
Ad_{g} \, \(\xi  + s {\hat k}\) =  g \xi g^{-1} \;+ \; {\hat k}
\( s +
\oint {d \varphi \o 2 \pi  } 
\tr \( g^{-1} \pa_\varphi g \xi \right)  \)
\lab{adj}
\ee
which in terms of the loop variable $\l= \exp (i\varphi)$
takes the following form
\be
Ad_{g} \,\( \xi  + s {\hat k}\) =  g \xi g^{-1} \;+ \; {\hat k}
\( s + {\rm Res}_{\l} 
\tr \( g^{-1} {d g \o d \l} \xi \right)  \)
\lab{adja}
\ee
In particular
\be
Ad_{\Theta} \,\( E^{(n)}\) =  \Theta E^{(n)} \Theta^{-1} \;+ \; {\hat k}
 {\rm Res}_{\l} \tr \( \Theta^{-1} {d \Theta \o d \l} E^{(n)}  \right)
= \Theta E^{(n)} \Theta^{-1} \;+ \; {\hat k} \cJ_n 
\lab{adj-bb}
\ee
Let $| 0 \rangle$ be the heighest weight state such that $X_{\geq 0} | 0 \rangle =0$
and $\langle 0 | X_{\leq 0}=0$ with 
$\langle 0 | {\hat k} | 0 \rangle=1$.
Then 
\be
\langle 0 | X_{+}  X_{-} | 0 \rangle = \omega ( X_{+},  X_{-})
=  {\rm Res}_{\l} \tr \( {d X_{+} \o d \l}  X_{-}\)
\lab{vacxx}
\ee
The $\tau$-function is defined by taking the centrally extended
the Riemann-Hilbert formula \rf{rh-def}
and multiplying from left and right by the vacuum states :
\be
\tau \( {\bf u}\)= \langle 0 | \exp \({\sum_{j=1}^{m+1} \sum_{n=1}^{\infty} 
E^{(n)}_{jj}u^{(n)}_j}\) \,  g \,
\exp \({-\sum_{j=1}^{m+1} \sum_{n=1}^{\infty} E^{(-n)}_{jj}u^{(-n)}_j} \)
 | 0 \rangle
\lab{tau-def}
\ee
similarly to the definition in \ct{Wilson:1993} for the AKNS hierarchy.
In the above definition the $\tau$-function depends on all the 
symmetry flows of the extended hierarchy.

Alternatively, for ${\hat M}$ being in a central extension ${\hat G}$ of $G$
over $M$ we can write the $\tau$-function as
$\tau \( {\bf u}\)= \langle 0 |\, {\hat M}\,| 0 \rangle=
\langle 0 |\, {\hat M_0}\,| 0 \rangle$
since the zero-grade term containg ${\hat k}$ resides in ${\hat M_0}$.

By setting all the negative flows $u^{(-n)}_j$ and positive flows
$u^{(-n)}_{j}, j \ne m+1$ to zero in \rf{tau-def}
we recover the standard $\tau$-function for the homogeneous $A_m$-hierarchy 
\ct{Wilson:1993}.
{}From \rf{tau-def} we find :
\be
{\pa \o \pa t_n }\tau \( {\bf u}\) = - \langle 0 | E^{(n)} 
{\hat \Theta^{-1}} ({\bf u})   {\hat M}  ({\bf u}) | 0 \rangle
= - \langle 0 | E^{(n)} 
{\hat \Theta^{-1}} ({\bf u})   | 0 \rangle \, \tau \( {\bf u}\)
\lab{pantau}
\ee
Using the property  $\langle 0 | X_{-}=0$ of the heighest weight state,
the last equation can be rewritten as
\be
{\pa \o \pa t_n }
\tau \( {\bf u}\) =  - \langle 0 | Ad_{\Theta} \,\( E^{(n)}\) 
| 0 \rangle \;\tau \( {\bf u}\) = - \cJ_n \; \tau \( {\bf u}\)
\lab{pantaua}
\ee
which confirms that the definition \rf{tau-def} reproduces the $\tau$-function
defined previously in \rf{jn-tau}.
More generally we find :
\be
{\pa \o \pa u^{(n)}_j}
\tau \( {\bf u}\) =  \langle 0 | Ad_{\Theta} \,\( E^{(n)}_{jj}\) 
| 0 \rangle \;\tau \( {\bf u}\) = - \cJ^{(n)}_j \; \tau \( {\bf u}\)
\lab{pantaub}
\ee

\subsection{ The Pseudo-differential Formalism of the Homogeneous ${\bf
A_m}$ Hierarchy and Construction of the M Matrix }
In this section we construct explicitly the $M( {\bf u}, \l)$
and $M^{-1}( {\bf u}, \l)$ matrices in terms of the objects appearing in the
pseudo-differential Lax calculus approach to the  homogeneous
$A_m$ hierarchy.
The matrix $M$ will be constructed as a expansion in positive powers of $\l$
which solves the spectral linear problem $L (M) =0$
in \rf{lmo}.
This will be accomplished by establishing relation between the algebraic
approach and the equivalent one based on 
the pseudo-differential Lax operator :
\begin{equation}
{\cal L} = {\partial_x} + \sum_{i=1}^{m} \Phi_i
{\partial_x}^{-1} \Psi_i 
\lab{lax-diff}
 \end{equation}
 and its inverse $\cL^{-1}$. Both operators can be represented
as ratios of two monic ordinary differential operators 
of order $m$ and $m+1$ \ct{krichev,Aratyn:2000dt,Aratyn:2000wr}:
\br
\cL &=& L_{m+1} L_m^{-1} = \pa_x + \sum_{i=1}^{m}  L_{m+1} ({{\phi}}_i) 
\partial_x^{-1}
{\psi}_i  \lab{cllmolm}\\
{\cal L}^{-1}  &=& L_{m} L_{m+1}^{-1} = \sum_{j=1}^{m+1}  L_m ({\bar {\phi}}_j) \partial_x^{-1}
{\bar \psi}_j
\lab{clilmlmo}
\er
Next, we provide few definitions and lemmas regarding the ordinary 
differential operators and Wronskians necessary 
for technical proofs to be shown in this section.
Let $L_m$ be a monic differential operator of order $m$ :
$L_m = \pa^m_x +u_{m-1} \pa_x^{m-1}+ \ldots + u_{1} \pa_x +u_0$
and  let $\p_i\;,\; i=1,{\ldots} ,m$ be
a basis for ${\rm Ker}\, L_m =\{\p_1, \ldots , \p_m\}$.
It follows that the action of $L_m$ is fully determined by $m$ elements of
its kernel :
\be
L_m (f) =  {W_{m} \lb \p_1, \ldots , \p_{m}, f \rb 
  \o W_{m} \lb \p_1, \ldots , \p_{m}\rb}
\lab{lmdef}
\ee
Here $W_{m} \lb \p_1, \ldots , \p_{m}\rb$ is a determinant 
of the Wronskian matrix:
\be
\cW_{m \times m} =\( \pa_x^{i-1} (\p_j)\)_{1 \leq i,j\leq m} 
\lab{wrdef}
\ee
which is nonsingular for the linearly independent functions $\p_1, \ldots , \p_m$.

In eq. \rf{cllmolm} we encounter $\psi_i,\, i=1,{\ldots} ,m$ which are
elements of the kernel of an adjoint operator 
$L_m^{\dagger}= (-1)^m \pa^m_x + (-1)^{m-1}\pa_x^{m-1}u_{m-1} + 
\ldots -  \pa_xu_{1} +u_0$:
\be
\psi_i = (-1)^{m+i} { W_{m-1} \lb \p_1, \ldots , {\widehat \p_{i}}, \ldots 
\p_m  \rb \o W_{m} \lb \p_1, \ldots , \p_{m}\rb } \quad ;\quad
\, i=1,\ldots ,m
\lab{psiipi}
\ee
with ${\widehat \p_{i}}$ being ommitted from the Wronskian.
Equation \rf{clilmlmo} contains elements
of $\left\{ {\bar {\phi}}_j \right\}_{j=1}^{m+1}$ and
$\left\{ {\bar {\psi}}_j \right\}_{j=1}^{m+1}$ in $Ker (L_{m+1})$ and 
$Ker(L_{m+1}^\dagger)$ connected with each other through a version
of \rf{psiipi}: 
\be
{\bar \psi}_j = (-1)^{m+1+j} { W_{m-1} \lb {\bar {\phi}}_1, \ldots ,
{\widehat {\bar \p}_{j}}, \ldots 
{\bar {\phi}}_{m+1}  \rb \o W_{m+1} \lb  {\bar {\phi}}_1, \ldots ,  
{\bar {\phi}}_{m}\rb } \quad ;\quad
\, j=1,\ldots ,m+1
\lab{barpsiipi}
\ee
Let $ \(\cW^{-1}_{m \times m}\)_{ij} \; i,j=0, {\ldots} , m  $ be 
the matrix elements of the inverse of the Wronskian matrix
$\cW_{m \times m}$. The following relations are then satisfied:
\be
\sum_{j=1}^{m} \(\cW^{-1}_{m \times m}\)_{ij} \p^{(j-1)}_k = \d_{i,k} \quad ; 
\quad \sum_{k=1}^{m} \p^{(j-1)}_k \(\cW^{-1}_{m \times m}\)_{kl} = \d_{j,l}
\lab{deltas}
\ee
By definition it holds that :
\be
\(\cW^{-1}_{m \times m}\)_{ij} = (-1)^{i+j} { \det_{(j,i)} \v\v 
\cW_{m \times m} \v\v \o W_{m} \lb \p_1, \ldots , \p_{m} \rb }
\lab{winverse}
\ee
where the determinant on the right hand side is the minor determinant 
obtained by extracting the $j$'th row and $i$'th column
from the Wronskian matrix $\cW_{m \times m}$ given in eq.\rf{wrdef}.

The following technical identity, which is valid for an arbitrary function
$\c$, follows directly from \rf{wrdef}-\rf{winverse}:
\be
\sum_{j=1}^{m} \(\cW^{-1}_{m \times m}\)_{ij} \c^{(j-1)}=
(-1)^{m+i} { W_{m} \lb \p_1, \ldots , {\widehat \p_{i}}, \ldots 
\p_m, \c \rb \o W_{m} \lb \p_1, \ldots , \p_{m}\rb } \quad ;\quad
i=1,\ldots ,m
\lab{winvc}
\ee

Due to the definition \rf{psiipi} the column $(\psi_1, \ldots ,\psi_m )^T$ 
agrees with the last column
in the inverse matrix $\cW^{-1}_{m \times m}$. 
As a consequence of this connection
we have a relation :
\be
\sum_{i=1}^m \p_i^{(k)} (t) \psi_i (t) = \d_{k,m-1} \quad{\rm for} \quad
k=0,1,\ldots ,m-1
\lab{wrona}
\ee
Let us introduce a notation :
\begin{equation}
 \Phi^{(-n)}_j = {\cal L}^{-n+1} \left( L_m ({\bar {\phi}}_j) \right),
 \;\;\; \Psi^{(-n)}_j = \({\cal L}^{\dagger}\)^{-n+1} \left( {\bar {\psi}}_j \right),
\quad j=1,{\ldots} ,m+1,\quad n=1,{\ldots} \, .
\lab{phineg}
\end{equation}
where ${\cal L}^{\dagger}= (L_m^{\dagger})^{-1} L_{m+1}^{\dagger}$.
For $n=1$ this reproduces the functions :
\be 
\Phi^{(-1)}_j= L_m ({\bar {\phi}}_j) \quad ,\quad
\Psi^{(-1)}_j =  {\bar {\psi}}_j ,
\quad j=1,{\ldots} ,m+1\, , 
\lab{phizeg}
\end{equation}
which satisfy relations 
\be
{\cal L} (\Phi^{(-1)}_j)=0  \quad ,\quad
{\cal L}^{\dagger} (\Psi^{(-1)}_j)=0 ,
\quad j=1,{\ldots} ,m+1 \,.
\lab{relphiz}
\ee

{}From \rf{wrona} we easily derive that
\be
{\rm Res}_{\pa_x}  {\cal L}^{-1} =  \sum_{j=1}^{m+1} 
L_m ({\bar {\phi}}_j) {\bar \psi}_j= \sum_{j=1}^{m+1} \Phi^{(-1)}_j  \Psi_{j}^{(-1)} =1
\lab{rescli}
\ee
Define now $F_j = \sum_{n=1}^{\infty} \lambda^{n-1} \Phi^{(-n)}_{j}$.
As pointed out in \ct{Aratyn:2000wr}, due to \rf{relphiz}
the $F_j $'s satisfy 
\begin{equation}
{\cal L} (F_j )  = \lambda  F_j  \,  
\lab{laxfj}
\end{equation}

The following definition appeared in \ct{Aratyn:2001cj} :
\begin{definition}
\lab{definition:fsphi}
Define the $(m+1) \times (m+1)$ matrix $M =(M_{ij})_{1\le i,j\le m+1}$ by
\begin{equation}
M_{m+1\,j} = F_j, \;\; \;   
M_{ij} = \partial_x^{-1}  \left( \Psi_i F_j  \right),\quad
i=1,{\ldots} ,m ,\;\;  j=1,{\ldots} , m+1
\lab{mdefs}
\end{equation}
\end{definition}
or in the matrix form :
\be
M \( {\bf u}, \l\)\,=\, \left(\begin{array}{ccc} 
\pa_x^{-1} \( \Psi_1 F_1  \) & \cdots & \pa_x^{-1} \( \Psi_1 F_{m+1}  \) \\
\vdots & \cdots & \vdots \\
\pa_x^{-1} \( \Psi_m F_1  \) & \cdots & \pa_x^{-1} \( \Psi_m F_{m+1}  \) \\
 F_1  & \cdots &  F_{m+1}  \end{array} \right)
\lab{mmma}
\ee
Due to \rf{mdefs} and \rf{laxfj} we find :
\ba
\partial_x M_{ij} = \Psi_i M_{m+1 \,j}, \quad i=1,{\ldots} ,m , \\
\left( \partial_x - \lambda \right) M_{m+1 \,j}
+ \sum_{i=1}^m  \Phi_{i}  M_{ij}  = 0,
\;\;\; j = 1, {\ldots}, m+1 \, .
\lab{mijlamb}
\ea
In terms of the matrix Lax operator
$L$ from \rf{matlaxdef} the above equation \rf{mijlamb}
is nothing but spectral problem  $L (M)=0$ \rf{lmo}.
This shows that the matrix $M$ constructed in
terms of the objects belonging to the pseudo-differential calculus
can be identified with the positive grade dressing matrix $M ({\bf u}, \l)$ 
of the extended Riemann-Hilbert problem.

Expanding now the $M({\bf u}, \lambda) $ as in
\be
M({\bf u}, \lambda)  = \sum_{i=1}^{\infty} M_i({\bf u})\, \l^i 
= M_0 + M_1 \l + {\ldots} 
\lab{mexp}
\ee
we find from the definition \rf{mdefs} 
for the zero-grade component of  $M ({\bf u}, \l)$ :
\be
(M_0)_{m+1\,j} = \Phi^{(-1)}_j, \;\; \;   
(M_0)_{ij} = \partial_x^{-1}  \left( \Psi_i \Phi^{(-1)}_j) \right),\quad
i=1,{\ldots} ,m ,\;\;  j=1,{\ldots} , m+1
\lab{mzdefs}
\ee
or in the matrix form :
\be
M_0 ({\bf u})\, = \,\left(\begin{array}{ccc} 
\pa_x^{-1} \( \Psi_1 \Phi^{(-1)}_1  \) & \cdots & 
\pa_x^{-1} \( \Psi_1 \Phi^{(-1)}_{m+1}  \) \\
\vdots & \cdots & \vdots \\
\pa_x^{-1} \( \Psi_m \Phi^{(-1)}_1  \) & \cdots & 
\pa_x^{-1} \( \Psi_m \Phi^{(-1)}_{m+1}  \) \\
 \Phi^{(-1)}_1  & \cdots &  \Phi^{(-1)}_{m+1}  \end{array} \right)
\lab{mmmaz}
\ee
We now propose the following matrix as an inverse of $M_0$ :
\be
(M_0)^{-1}_{j\, m+1} = \Psi^{(-1)}_j, \;\; \;   
(M_0)_{ji}^{-1}  = \partial_x^{-1}  \left( \Phi_i \Psi^{(-1)}_j  \right),\quad
i=1,{\ldots} ,m ,\;\;  j=1,{\ldots} , m+1
\lab{mzinvdefs}
\ee
which in the matrix form is given by :
\be
M_0^{-1}  ({\bf u}) \, = \,\left(\begin{array}{cccc} 
\pa_x^{-1} \( \Phi_1 \Psi^{(-1)}_1  \) & \cdots & 
\pa_x^{-1} \( \Phi_m \Psi^{(-1)}_{1}  \)
& \Psi^{(-1)}_{1} \\
\vdots & \cdots & \cdots &\vdots \\
\pa_x^{-1} \( \Psi_1 \Psi^{(-1)}_{m+1}  \) & \cdots & 
\pa_x^{-1} \( \Psi_m \Psi^{(-1)}_{m+1}  \) &
  \Psi^{(-1)}_{m+1}  \end{array} \right)
\lab{mmmazinv}
\ee
We now provide a proof of this statement. We first notice that
$\sum_{j=1}^{m+1} (M_0)_{m+1\,j} (M_0)^{-1}_{j\, m+1} =1$ due to \rf{rescli}.
Furthermore, $\sum_{j=1}^{m+1} (M_0)_{m+1\,j}(M_0)_{ji}^{-1} =0$
and $\sum_{j=1}^{m+1} (M_0)_{m+1\,j}(M_0)_{ji}^{-1} =0$
as follows from relations
$\cL^{-1}(\P_i)=0,  \({\cal L}^{\dagger}\)^{-1} (\Psi_i)=0, i=1,{\ldots} ,m$.

It remains to show the $m \times m$ identities :
\be 
\sum_{j=1}^{m+1} (M_0)_{i\,j} (M_0)^{-1}_{j\, k}
= \sum_{j=1}^{m+1} \pa_x^{-1}  \left( \Psi_i \Phi^{(-1)}_j\right)
\partial_x^{-1}  \( \Phi_k \Psi^{(-1)}_j  \right) =\d_{ik}
\lab{mzijmz}
\ee
Consider first a factor : 
\ba
\pa_x^{-1}  \( \Phi_k \Psi^{(-1)}_j  \)=
\pa_x^{-1}  \( L_{m+1} (\p_k)  {\bar \psi}_j \) 
\\
= (-1)^{m+1+j} \pa_x^{-1} 
\( W \lb {\bar \p}_1, {\ldots} ,{\bar \p}_{m+1}, \p_k \rb
 W \lb {\bar \p}_1 ,{\ldots} ,{\widehat {\bar \p}_j}
 ,{\ldots} ,{\bar \p}_{m+1} \rb /
 W^2 \lb {\bar \p}_1, {\ldots} ,{\bar \p}_{m+1}\rb \)
\\
=  (-1)^{m+1+j} 
W \lb {\bar \p}_1 ,{\ldots} ,{\widehat {\bar \p}_j}
,{\ldots} ,{\bar \p}_{m+1}, \p_k  \rb / 
W \lb {\bar \p}_1, {\ldots} ,{\bar \p}_{m+1}\rb 
\lab{idwrona}
\ea
where use was made of the Jacobi theorem :
\be
 \pa_x \( { W_{k-1 } (f) \over W_{k}} \) = {W_{k} \(f\) W_{k-1 }
\over W_{k}^2}
\lab{jac-pr}
\ee
involving Wronskians :
\be
W_k \equiv W_k \lb g_1, \ldots ,g_k \rb
\;\;\; ;\;\;\; 
W_{k-1} \(f \)\equiv W_{k} \lb g_1, \ldots ,g_{k-1}, f\rb.
\ee
for some arbitrary functions $g_1, \ldots ,g_k$.

Using the same technique we also find that
\ba
\pa_x^{-1}  \( \Psi_i \Phi^{(-1)}_j) \) = 
\pa_x^{-1}  \( \psi_i L_m ({\bar \p}_j \)  \\
=  (-1)^{m+j} 
W \lb {\p}_1 ,{\ldots} ,{\widehat {\p}_j}
,{\ldots} ,{\p}_{m}, {\bar \p}_j  \rb / 
W \lb {\p}_1, {\ldots} ,{\p}_{m} \rb 
\lab{idwronb}
\ea
Therefore using relation \rf{winvc} the quantity in \rf{mzijmz} becomes
\br
\sum_{j=1}^{m+1} \pa_x^{-1}  \left( \Psi_i \Phi^{(-1)}_j \right)
\partial_x^{-1}  \( \Phi_k \Psi^{(-1)}_j  \right) &=&
\sum_{j=1}^{m+1} \sum_{s=1}^{m} 
\(\cW^{-1}_{m \times m}\)_{is} {\bar \p}^{(s-1)}_j
\sum_{l=1}^{m+1} \p^{(l-1)}_k \(\cW^{-1}_{m+1 \times m+1}\)_{jl} \nonu\\
&=& \sum_{s=1}^{m} \sum_{l=1}^{m+1} \d_{sl}
\(\cW^{-1}_{m \times m}\)_{is} \p^{(l-1)}_k
= \d_{ik}
\lab{ndeltas}
\er
as required for the proof of $M_0^{-1}$ being an inverse of $M_0$.

Next, define $G_{j} (\l)=
\sum_{n=1}^{\infty} \l^{n-1} \Psi_{j}^{(-n)}$ which is the
solution of the conjugated spectral
problem $\cL^{\dagger} (G_{j} (\l)) = \l G_{j} (\l)$.

We now construct the inverse of the $M$ matrix  which  we denoted by
$M^{-1} =(M^{-1}_{ij})_{1\le i,j\le m+1}$.
In view of  $L (M)=0$ in \rf{lmo}
the matrix elements of $M^{-1}$ must satisfy :
\br
\partial_x M_{ji}^{-1} &=& \Phi_i M_{j \, m+1}^{-1}, 
\quad i=1,{\ldots} ,m ,\nonu \\
\left( \partial_x + \lambda \right) M_{j \, m+1}^{-1}
 &+& \sum_{i=1}^m  \Psi_{i}  M_{ji}^{-1}  = 0,
\;\;\; j = 1, {\ldots}, m+1 \, .
\lab{invmijlamb}
\er
or in the matrix notation :
\be 
\partial_x M^{-1} + M^{-1} \( E^{(1)}_{m+1 \, m+1} - A\) =0
\lab{minvmat}
\ee
This provides a recurrence relation which allows calculation of
terms of higher grade in $M^{-1}$ from $ M^{-1}_0$.
The result in terms of $G_{j} (\l)$ gives 
the $(m+1) \times (m+1)$ matrix $M^{-1}$ given by
\begin{equation}
M^{-1}_{j\, m+1} = G_j, \;\; \;   
M_{ji}^{-1}  = \partial_x^{-1}  \left( \Phi_i G_{j}  \right),\quad
i=1,{\ldots} ,m ,\;\;  j=1,{\ldots} , m+1
\lab{mnvdefs}
\end{equation}
Observe, that 
\be
\sum_{j=1}^{m+1} F_j \pa_x^{-1} G_j = 
\sum_{j=1}^{m+1} \sum_{n=0}^{\infty} \sum_{l=0}^{\infty} 
\l^n \Phi^{(-n+l-1)}_j \pa_x^{-1} \Psi_{j}^{(-l-1)}
= \sum_{n=0}^{\infty} \l^n \cL^{-(n+1)}
\lab{fdgcln}
\ee
where $\cL^{-n} = \sum_{j=1}^{m+1} \sum_{l=1}^{n}
\Phi^{(-n+l-1)}_j \pa_x^{-1} \Psi_{j}^{(-l)}$ generalizes
expression for $ {\cal L}^{-1}$.
The result \rf{mnvdefs} regarding the $M^{-1}$ is further supported
by the following identities:
\be
\sum_{j=1}^{m+1} F_j  G_j =
{\rm Res}_{\pa_x} \(\sum_{j=1}^{m+1} F_j \pa_x^{-1} G_j\)
= {\rm Res}_{\pa_x} {\cal L}^{-1} 
 =1
 \lab{fgone}
\ee
which follows from \rf{fdgcln} and 
\be
\sum_{j=1}^{m+1} F_j \pa_x^{-1} \( G_j \Phi_{i}\) = 0
\;\; , \;\; \sum_{j=1}^{m+1} G_j \pa_x^{-1} \( F_j \Psi_{i}\) = 0
, \quad i=1,{\ldots} ,m
\lab{fgpp}
\ee
following from ${\cal L}^{-1} (\Phi_i) = 0 $ and
${\cal L}^{\dagger\, -1} (\Psi_i) = 0 $ valid for $i=1,{\ldots} ,m$.

Here we illustrate the above construction for $m=1$ with the AKNS Lax 
operator $\cL = D -r D^{-1} q$ defining 
the spectral problem $\cL (\psi) = \l \psi$.
The self-commuting isospectral flows ($n>0$): 
$ \pa_n r = B_n (r)$ and $\pa_n q = - B_n^\dagger (q)$ with $B_n = 
(\cL^n)_{+}$ belong to the positive part of the AKNS hierarchy.
$\cL$ can be described as a ratio of two ordinary monic
differential operators as
$\cL = L_2 L_1^{-1}$, where $ L_1,L_2$ denote monic operators
$ L_1 = (D+ \vp_1^{\pr}+ \vp_2^{\pr})$ and 
$ L_2 = (D+ \vp_1^{\pr}) (D+ \vp_2^{\pr})$
of, respectively, order $1$ and $2$.
A monic differential operator $L_2$ is fully characterized
by elements of its kernel, $\p_1 = \exp (- \vp_2)$ and
$\p_2 = \exp (- \vp_2)\int^x \exp ( \vp_2-\vp_1)$.
Its inverse $L_2^{-1}$, is given by
$L_2^{-1} = \sum_{\a=1}^2 \p_{\a} D^{-1} \psi_{\a}$, where
$\psi_1 = - \exp ( \vp_1)\int^x \exp ( \vp_2-\vp_1)$
and $\psi_2 =  \exp ( \vp_1)$ are kernel elements of the conjugated
operator $L_2^{\dagger}= (-D+ \vp_2^{\pr})(-D+ \vp_1^{\pr}) $, see \ct{UIC-97} and
references therein.
In this notation, $ \cL = D + L_2 (\exp ( -\vp_1-\vp_2))\,D^{-1} \exp (
\vp_1+\vp_2)$ and accordingly:
\be
q= -\exp (\vp_1+\vp_2) \; \; ; \; \; r= -\( \vp_1^{\pr \pr} - \vp_1^{\pr} 
\vp_2^{\pr} \) \exp (-\vp_1-\vp_2)
\lab{qr-phis}
\ee
Similarly, the inverse of $\cL$ is also given as a ratio of differential
operators
$\cL^{-1} = L_1  L_2^{-1}=  \sum_{\a=1}^2 L_1(\p_{\a}) D^{-1} \psi_{\a}$.
The functions $\P_{\a}^{(-1)} \equiv  L_1(\p_{\a})$ and
$ \Psi_{\a}^{(-1)} \equiv \psi_{\a}$ satisfy the same flow equations
as $r$ and $q$ with respect to the positive flows of the AKNS hierarchy.

To facilitate comparison with the subsection 4.3
we introduce variables $R,u,v$ 
\be
R=\vp_1 \;\; ; \; \; u = e^{\vp_1}\,\int^x e^{\vp_2-\vp_1}  \;\; ; \; \;
v = \vp_1^{\pr}\,e^{-\vp_2}
\lab{ident}
\ee
in terms of $\varphi_1 $ and $\varphi_2$ \ct{Aratyn:2000dt}.

{}Based on expressions \rf{mzdefs}-\rf{mzinvdefs} obtained by the 
pseudo-differential approach we write $M_0$ and its inverse as :
 \br
M_0 =  \fourmat{\pa^{-1} (\Phi \Psi_1^{(-1)} )}{ \Psi_1^{(-1)} }
{\pa^{-1} (\Phi \Psi_2^{(-1)} ) }{ \Psi_2^{(-1)} }, \quad 
M_0^{-1} =  \fourmat{\pa^{-1} (\Psi \Phi_1^{(-1)} )}{\pa^{-1} (\Psi \Phi_2^{(-1)} 
) }{ \Phi_1^{(-1)} }{ \Phi_2^{(-1)} }
\lab{mzeropseudo}
\er
and within the constrained AKNS we obtain
\br
\Phi_1^{(-1)} &=& \varphi_1^{\pr} e^{-\varphi_2} = -u = \Psi_1^{(-1)} = - e^{\varphi_1}\int^{x} e^{\varphi_2 -
\varphi_1}\nonu \\
\Phi_2^{(-1)} &=& \varphi_1^{\pr} e^{-\varphi_2}\int^{x} e^{\varphi_2 -
\varphi_1} + e^{-\varphi_1} = \Delta e^{-R} = \Psi_2^{(-1)} = e^{R}
\er
Recalling, that for the CKP hierarchy \cite{Date:1981rb,Aratyn:2001cj}
\br
\Psi = \Phi = -q = r
\er
 and using \rf{qr-phis} , we obtain
 \br
\pa^{-1} ( \Psi \Phi_1^{(-1)} ) = e^{R}, \quad \pa^{-1} ( \Psi \Phi_2^{(-1)} ) = u
\lab{pppaa}
\er
In a similar way, we find 
\br
\pa^{-1} ( \Phi \Psi_1^{(-1)} ) = e^{R}, \quad \pa^{-1} ( \Phi \Psi_2^{(-1)} ) = -u
\lab{pppab}
\er

\sect{The Multidimensional Toda Model
 and the Cecotti-Vafa Equations}

\subsection{The Multidimensional Toda Model}

Introduce a notation :
\be
\pa_j \equiv {\pa \o \pa u^{(1)}_j } \quad; \quad 
\pa_{-j} \equiv {\pa \o \pa u^{(-1)}_j } \; , \quad
\lab{pajdef}
\ee
With this notation the relevant part of the flow equations
\rf{uthpos}-\rf{ummmin} takes a form 
\ba
\pa_j \Theta ({\bf u}, \lambda) 
\;=\; - \left(\Theta E^{(1)}_{jj} \Theta^{-1} 
\right)_{-} \Theta ({\bf u}, \lambda) 
\\
\pa_j M ({\bf u}, \lambda) 
\;=\;  \left(\Theta E^{(1)}_{jj} \Theta^{-1} 
\right)_{+} M ({\bf u}, \lambda) 
\lab{pajthm} 
\ea
and
\ba
\pa_{-j} \Theta ({\bf u}, \lambda) 
\;=\;  \left( M E^{(-1)}_{jj} M^{-1} 
\right)_{-} \Theta ({\bf u}, \lambda) 
\\
\pa_{-j}  M ({\bf u}, \lambda) 
\;=\; -  \left( M E^{(-1)}_{jj} M^{-1} 
\right)_{+} M ({\bf u}, \lambda) 
\lab{pamjthm}
\ea 
The last equation in \rf{pamjthm} can also be rewritten as
\br
\pa_{-j}  M ({\bf u}, \lambda) &=& - \( M E^{(-1)}_{jj} M^{-1} - 
( M E^{(-1)}_{jj} M^{-1} )_{-}\)  M \nonu \\
&=& -  M E^{(-1)}_{jj} + M_0 E^{(-1)}_{jj} M_0^{-1}   M
\lab{pamjthma}
\er
Projecting \rf{pamjthma} on the zero grade 
and recalling expansion in \rf{mexp} we find 
\be
\pa_{-j}  M_0 = -  M_1  E_{jj} + M_0 E_{jj} M_0^{-1} M_1 
\lab{pamjmz}
\ee
which can be cast in the following form :
\be
M_0^{-1} \, \pa_{-j}  M_0 = \lbrack \, E_{jj} \,, \, M_0^{-1} M_1 \,
\rbrack
\lab{pamjmza}
\ee
Similarly, by projecting the second equation in \rf{pajthm} on 
grades zero and one, we find :
\br
\pa_{j}  M_0 &=& \lbrack \, \theta^{(-1)}\,, \,  E_{jj} \rbrack \, M_0
,\;\;\;\;\;\; j =1,{\ldots} ,m+1
\lab{paimz}\\
\pa_{i}  M_1 &=& E_{jj}  M_0 +
\lbrack \, \theta^{(-1)}\,, \,  E_{jj} \rbrack \, M_1
\lab{paimo}
\er
Using \rf{pamjmz} and \rf{paimz} we can cast the flow equations
\rf{pajthm} and \rf{pamjthm} in a way which reveals a symmetry between
the negative and positive flows and the dressing matrices of the positive
and negative gradations.
Replacing $M ({\bf u}, \lambda)$ by $M_0^{-1} M ({\bf u}, \lambda)$ in 
\rf{pajthm} and \rf{pamjthm} we find :
\ba
\pa_{-j} \Theta ({\bf u}, \lambda) 
\;=\;  \left( M_0 E^{(-1)}_{jj} M^{-1}_0 
\right) \Theta ({\bf u}, \lambda) 
\\
\pa_{j}  \( M_0^{-1} M ({\bf u}, \lambda) \)
\;=\; \left( M_0^{-1} E^{(1)}_{jj} M_0 
\right)\( M_0^{-1} M ({\bf u}, \lambda) \) 
\lab{papmjthm}
\ea 
and 
\ba
\pa_{j} \Theta ({\bf u}, \lambda) 
\;=\; -  \left(\Theta E^{(1)}_{jj} \Theta^{-1} 
\right)_{-} \Theta ({\bf u}, \lambda) 
\\
\pa_{-j}  \( M_0^{-1} M ({\bf u}, \lambda) \)
\;=\; - \left(  M_0^{-1} M E^{(-1)}_{jj} (M^{-1}_0 M)^{-1}
\right)_{>0} \( M_0^{-1} M ({\bf u}, \lambda) \) 
\lab{pampjthm}
\ea
Equations \rf{papmjthm} and \rf{pampjthm} exhibit invariance under
simultaneous interchanges : $\pa_j \leftrightarrow \pa_{-j} $,
$\Theta ({\bf u}, \lambda) \leftrightarrow  M_0^{-1} M ({\bf u}, \lambda) $,
$M_0 \leftrightarrow  M_0^{-1} $ and of the positive and negative grades.
This type of symetry will be responsible for appearance of the complex like
structure  among  the Cecotti-Vafa equations we will derive later
in this section.

Applying now $\pa_i $ on equation \rf{pamjmza} and using relations 
\rf{paimz} and \rf{paimo} we obtain
\be
\pa_i \(M_0^{-1} \, \pa_{-j}  M_0\) 
= \Bigl\lbrack \, E_{jj} \,, \, -M_0^{-1} 
\lbrack \, \theta^{(-1)}\,, \,  E_{ii} \rbrack \, M_1 \,
+M_0^{-1} \( E_{ii} M_0 + 
\lbrack \, \theta^{(-1)}\,, \,  E_{ii} \rbrack \, M_1 \) \,
\Bigr\rbrack 
\lab{paipamjmz}
\ee
which after the cancellation of two identical terms with opposite signs
results in :
\be
\pa_i \(M_0^{-1} \, \pa_{-j}  M_0\) 
= \lbrack \, E_{jj} \,, \, M_0^{-1} E_{ii} M_0 \,
\rbrack, \;\;\;\; i,j =1,{\ldots} ,m+1
\lab{paipamjmza}
\ee
By multiplying both sides of \rf{paipamjmza} by $M_0$ from the left and
$M_0^{-1}$ from the right we obtain an equivalent expression :
\be
\pa_{-j} \( \pa_{i}  M_0 \, M_0^{-1}\) 
= \lbrack \, M_0 E_{jj} M_0^{-1} \,, \,  E_{ii} \,
\rbrack , \;\;\;\; i,j =1,{\ldots} ,m+1
\lab{pamjpaimza}
\ee
which can be rewritten as a Toda zero-curvature equation:
\be
\lbrack \, \pa_{-j} - M_0  E^{(-1)}_{jj}M^{-1}_0 
\, , \,  \pa_i - E^{(1)}_{ii} -\(  \pa_{i}  M_0 \) M^{-1}_0
\, \rbrack =0\, .
\lab{toda-ov-id}
\ee
Consider next:
\br 
\pa_{-i} \(M_0\, E_{jj}\,   M_0^{-1} \) &=&
\(-M_1 E_{ii} + M_0 E_{ii} M_0^{-1} M_1 \) 
E_{jj}  M_0^{-1} \nonu \\
&+& M_0 E_{jj}\( M_0^{-1} M_1 E_{ii} M_0^{-1}
-E_{ii} M_0^{-1} M_1M_0^{-1} \) \, .
\lab{pamipamj}
\er
For $i\ne j$ it holds that $E_{ii} E_{jj} =0$.
Using the $i \leftrightarrow j$
symmetry exhibited by two remaining terms on the right hand side of 
\rf{pamipamj} we obtain 
\be
\pa_{-i} \(M_0 \, E_{jj}\,   M_0^{-1} \) 
= \pa_{-j} \(M_0 \, E_{ii}\,   M_0^{-1} \)\, .
\lab{pamipamja}
\ee
In the same way we also obtain :
\be
\pa_{i} \(M_0^{-1} \, E_{jj}\,   M_0\) 
= \pa_{j} \(M_0^{-1} \, E_{ii}\,   M_0\) \, .
\lab{paipaja}
\ee
On basis of relations \rf{paipamjmza}, \rf{pamipamja} and \rf{paipaja}
we recognize that $M_0$ satisfies the multidimensional Toda model
\ct{Razumov:1996me,Ferreira:1999px}.

\subsection{Orthogonal Reduction of the 
  ${\bf \widehat{gl} (m+1, \IC)}$-Hierarchy} 

Consider ${\cgh} = {\widehat sl}(m+1)$ with Cartan subalgebra generators
$H_a^{(n)}$ and step operators $E^{(n)}_{ij}$ with $n \in \IZ$ and $i\ne j$.
Next, define the extended automorphism $\sigma$, 
such that
\br
\sigma \( H_a^{(n)}\) &=& - (-1)^n H_a^{(n)} \quad a=1,{\ldots} , m
\lab{hma}\\
\sigma \( E^{(n)}_{ij}\) &=& - (-1)^n E^{(n)}_{ji} \quad i\ne j=
1,{\ldots} , m+1
\lab{seij}
\er
The $\sigma$ automorphism agrees, for $n=0$, with the well-known
automorphism defining the symmetric space $sl (m+1) /so(m+1)$ \ct{helgason}.
The combinations
\be
H_a^{(2n+1)}, \, a=1,{\ldots} , m, \; E^{(2n)}_{ij} - E^{(2n)}_{ji}, \;
E^{(2n+1)}_{ij} + E^{(2n+1)}_{ji}\quad i\ne j=
1,{\ldots} , m+1\;,\; n \in \IZ
\lab{combins}
\ee
generate the subalgebra of ${\widehat sl}(m+1)$ invariant under 
automorphism $\sigma$.
Let 
\be
E_{\sigma} = \h \( E^{(1)} + \sigma ( E^{(1)}) \)= \mu_m \cdot H^{(1)} 
\lab{rede}
\ee
and consider the kernel of ${\rm ad} (E_{\sigma} ) $
within the subalgebra of ${\widehat sl}(m+1)$ invariant under
automorphism $\sigma$. Such kernel is generated by even combinations
from \rf{combins} within ${\widehat sl}(m)\otimes {\hat u} (1)$.
The image of ${\rm ad} (E_{\sigma} ) $ 
is generated  by the following combinations :
\be
E^{(2n)}_{m+1\, i} - E^{(2n)}_{i\,m+1}, \;
E^{(2n+1)}_{m+1\,i} + E^{(2n+1)}_{i\, m+1}\quad i=
1,{\ldots} , m\;,\;\;\; n \in \IZ
\lab{combinsa}
\ee
The corresponding reduced potential lies in the zero-grade
subspace spanned by \rf{combinsa}, i.e.
\be
A_{\sigma} = \sum_{i=1}^m \Phi_i \( E^{(0)}_{m+1\, i} - E^{(0)}_{i\,m+1} \)
\lab{asigm}
\ee
with the property that
\be
A_{\sigma} = - A_{\sigma}^T
\lab{prop}
\ee
The loop group generalization of the automorphism in 
\rf{hma}-\rf{seij} has the following form \ct{vandeLeur:2000gk} :
\be
\sigma \( X ( \l ) \)  = \( \( X( -\l ) \)^T \)^{-1}\quad;\quad
X \in G= \widehat{GL} (m+1)
\lab{lgsauto}
\ee
One notices that the evolution equations \rf{uthpos},\rf{ummpos},
\rf{uthneg} and \rf{ummmin} 
are invariant under the automorphism  $\sigma$ defined in
\rf{lgsauto} for $n$ being an odd integer.
As an illustration we consider equation \rf{uthpos}
and find that the flows for the $\sigma$ transformed $\Theta$ matrix become :
\be
 \frac{\partial}{\partial u^{(n)}_j}
\s\( \Theta\) ( {\bf u}, \lambda )  = 
(-1)^n \(\s \(\Th\)  ( {\bf u}, \lambda )  E^{(n)}_{jj}  
\s \( \Theta^{-1}\) ( {\bf u}, \lambda )\)_{-}
\s\( \Theta\)  ( {\bf u}, \lambda ) 
\lab{Th-flowsa-tran}
\ee
Equation \rf{Th-flowsa-tran} agrees with \rf{uthpos}
for $(-1)^n =-1$ which shows the desired result.

Accordingly, we define the integrable sub-hierarchy by constraining the
dressing matrices $\Theta ({\bf u}, \l )$ and
$ M ({\bf u}, \l )$  to be the fixed points of the
loop group  automorphism $\sigma $ \rf{lgsauto} :
\br
\Theta^{-1} ( {\bf u}, \lambda ) &=&     \Theta^T ( {\bf u}, -\lambda )
\lab{red-theta}\\
M^{-1} ( {\bf u}, \lambda ) &=&    M^T ( {\bf u}, -\lambda )
\lab{red-thetb}
\er
with $\Theta ({\bf u}, \l )$ and
$ M ({\bf u}, \l )$ depending only on odd coordinates ${\bf u}$:
( $u^{(n)}_j = u^{(2k+1)}_j$).
{}From our discussion above it is clear that the odd flows of the reduced 
sub-hierarchy will preserve the conditions \rf{red-theta}
and \rf{red-thetb}.

The fixed points of the automorphism $\sigma$ form a subgroup of
 $G= \widehat{GL} (m+1)$, called a twisted loop group of $GL (m+1)$.
In reference \ct{vandeLeur:2000gk}, the twisted loop group of $GL (n)$,
in the context of $n$-component KP hierarchy, was used to find solutions
of the Darboux-Egoroff system of PDE's.

Note, that from \rf{red-thetb} we derive the additional 
orthogonality constraint on $M_0$ :
\be 
M_0^T = M_0^{-1}
\lab{ortho-cond}
\ee
For the first term $\theta^{(-1)}$ of expansion of
$\Theta = 1 + \theta^{(-1)} + O(\l^{-2})$ the constraint
\rf{red-theta} implies that $\theta^{(-1)} = \theta^{(-1)\, T}$.
Hence the reduction based on \rf{red-theta} 
imposes $A^T =-A$ as in \rf{prop}.

Let us now return to the extended Riemann-Hilbert problem \rf{rh-def}
and restrict it to the reduced model with only odd-flows and 
with constraints \rf{red-theta}-\rf{red-thetb}.
By writing the extended Riemann-Hilbert problems \rf{rh-def}
for both $\Theta^{-1} ( {\bf u}, \lambda )$ and 
$\Theta^T ( {\bf u}, -\lambda )$  with the constraint \rf{red-theta}
imposed we obtain :
\br
M^T ( {\bf u}, -\l ) M ( {\bf u}, \l) &=&
\exp \({\sum_{j=1}^{m+1} \sum_{k=0}^{\infty} 
E^{(-2k-1)}_{jj}u^{(-2k-1)}_j} \) g^T ( -\l) g (\l) 
\nonu \\
&\times&
\exp \(-{\sum_{j=1}^{m+1} \sum_{k=0}^{\infty} 
E^{(-2k-1)}_{jj}u^{(-2k-1)}_j} \) 
\lab{mtm}
\er
Due to \rf{red-thetb} we see that the necessary condition for
the $g$ group element is :
\be
g^T ( -\l) g (\l) = I
\lab{red-thetg}
\ee
i.e. $ g (\l) $ is a fixed point of the automorphism $\sigma$.
Alternatively, we can derive the reduced sub-hierarchy
from the Riemann-Hilbert problem with odd flows
defined on the twisted loop group of $GL (m+1)$.

The tau function for this sub-hierarchy becomes :
\be
\tau \( {\bf u}\)= \langle 0 | \exp \({\sum_{j=1}^{m+1} \sum_{k=0}^{\infty} 
E^{(2k+1)}_{jj}u^{(2k+1)}_j}\) \,  g (\l)\,
\exp \({-\sum_{j=1}^{m+1} \sum_{k=0}^{\infty} E^{(-2k-1)}_{jj}u^{(-2k-1)}_j} \)
 | 0 \rangle
\lab{tau-def-red}
\ee
and can be obtained from the original $\tau$-function in \rf{tau-def} by
putting $u^{(\pm 2k)}_j=0$ and enforcing on $g (\l)$ the 
condition \rf{red-thetg}.  The model is therefore embedded in
the CKP hierarchy \ct{Date:1981rb,Aratyn:2001cj}.

\subsection{Cecotti-Vafa  Equations}

Next, observe that from \rf{pamjmz} it follows :
\be
\sum_{j=1}^{m+1} \pa_{-j} M_0 =  0 \, .
\lab{sumpajmz}
\ee
Using \rf{pamjthma} and \rf{pamjmz} we find : 
\be
\pa_{-j}  \(M_0^{-1} M_1\) = \lbrack \, M_0^{-1} M_1 \, , \,   E_{jj}
\, \rbrack \,M_0^{-1} M_1  + \lbrack \, E_{jj} \, , \, M_0^{-1} M_2
\, \rbrack 
\lab{pamjmzmo}
\ee
which leads to :
\be
\sum_{j=1}^{m+1} \pa_{-j} (M_0^{-1} M_1) = 0
\lab{sumpajmzmo}
\ee
Moreover, from \rf{pamjmz} and \rf{pamjmzmo} we get for the matrix 
components $(M_0)_{ik}$ and $(M_0^{-1} M_1)_{ik}$ :
\br
\pa_{-j}  (M_0)_{ik} &=& (M_0)_{ij} (M_0^{-1} M_1)_{jk}
\quad ; \quad j \ne k
\lab{pamjmzik}\\
\pa_{-j}  (M_0^{-1} M_1)_{ik} &=&  (M_0^{-1} M_1)_{ij} (M_0^{-1} M_1)_{jk} 
\quad ; \quad i\, ,j \, ,k\, \;{\rm distinct}
\lab{pamjmzmoik}
\er
{}From
\be
\pa_{j}  \(M_0^{-1} M_1\) = M_0^{-1} E_{jj} M_0
\lab{paim0invm1}
\ee
we get
\be
\pa_{j}  (M_0^{-1} M_1)_{ik} =  (M_0^{-1})_{ij} (M_0)_{jk}
\lab{paim0invm1jk}
\ee
Similarly, from \rf{paimz} we find :
\be
\pa_{j}  (M_0)_{ik} =  \( \theta^{(-1)}\)_{ij} (M_0)_{jk} 
\quad ; \quad i\ne j 
\lab{pajmzik}
\ee
For $ \theta^{(-1)}$ we find from \rf{pajthm} :
\be
\pa_{j}  \theta^{(-1)}  = 
\lbrack \,   E_{jj}   \, , \, \theta^{(-2)}  
\, \rbrack \  + \lbrack \, \theta^{(-1)} \, , \,  E_{jj} 
\, \rbrack \, \theta^{(-1)}
\lab{pajthm1}
\ee
from which it follows that :
\be
\pa_{j}  (\theta^{(-1)})_{ik} = \( \theta^{(-1)}\)_{ij} 
(\theta^{(-1)})_{jk} 
\quad ; \quad i\, ,j \, ,k\, \;{\rm distinct}
\lab{pajthik}
\ee
Also 
\be
\pa_{-j}  \theta^{(-1)}  = M_0 \,E_{jj} \, M_0^{-1}
\lab{pamjthmz}
\ee
gives
\be
\pa_{-j}  (\theta^{(-1)})_{ik} =  (M_0)_{ij}\, (M_0^{-1})_{jk} 
\lab{pamjthmzik}
\ee

Consider now the reduced case with the orthogonal matrix:
$M_0 =(m_{ij})_{1\le i,j\le m+1}$.
For simplicity we introduce a notation
$M_0^{-1} M_1 = {\overline \cB}= ({\bar \beta}_{ij})_{1\le i,j\le m+1}$. Then from 
\rf{pamjmza} it follows that 
\be
M_0^T \pa_{-j} M_0  = \lbrack \,   E_{jj}   \, , \, {\overline \cB}\, \rbrack 
\lab{bbeta}
\ee
for all $j$ such $1 \leq j \leq m+1$. Transposing both sides of
the matrix relation \rf{bbeta} we find that the matrix
$M_0^{-1} M_1={\overline \cB}$
is symmetric (${\overline \cB}^T={\overline \cB}$) for the matrix $M_0$ satisfying the orthogonality
condition \rf{ortho-cond}. 

We can summarize our results \rf{paipamjmza},\rf{pamipamja} and \rf{paipaja}
in the reduced case as :
\br
\pa_i \(M_0^T \pa_{-j} M_0 \) &=& 
\lbrack \,   E_{jj}   \, , \,  M_0^T  E_{ii} M_0 \rbrack 
\lab{paipamjbet} \\
\pa_{-i} \(M_0 \, E_{jj}\,   M_0^T\) &=& 
 \pa_{-j} \(M_0 \, E_{ii}\,   M_0^T\)
\lab{pamipamjb}\\
\pa_{i} \(M_0^T  \, E_{jj}\,   M_0\) &=& 
 \pa_{j} \(M_0^T \, E_{ii}\,   M_0 \)
\lab{paipajbet}
\er
These equations have been derived by Cecotti and Vafa in 
\ct{Cecotti:1991me,Cecotti:1993vy}.
Imposing the hermiticity condition $M^{T}_0 =M_0^{*}$ or $M_0=M_0^{\dagger}$, 
we find that the equations \rf{paipamjbet}-\rf{paipajbet}
are invariant under complex conjugation ${}^{*}$ which takes
$\pa_j$ to $\pa_{-j}$ and viceversa.

The ${\overline \cB}$-matrix elements satisfy :
\br
\pa_{-j}  {\bar \beta}_{ik} &=&  {\bar \beta}_{ij} {\bar \beta}_{jk} 
\quad ; \quad i\, ,j \, ,k\, \;{\rm distinct}
\lab{pamjmzmoikb}\\
\sum_{j=1}^{m+1} \pa_{-j} {\bar \beta}_{ik} &=& 0 
\lab{summjB}\\
\pa_{j}  {\bar \beta}_{ik} &=&  m_{ji} m_{jk}
\lab{paim0invm1jkb}
\er
as follows from relations \rf{sumpajmzmo}, \rf{pamjmzmoik} and
\rf{paim0invm1jk}.
The first two equations \rf{pamjmzmoikb} and \rf{summjB} for 
the symmetric ${\overline \cB}$ matrix are characteristic for the Egoroff metric.

For the derivatives of matrix elements $m_{ij}$ we find 
from \rf{pamjmz}, \rf{sumpajmz} and \rf{pamjmzik} :
\br
\pa_{-j}  m_{ik} &=& m_{ij} {\bar \beta}_{jk}
\quad ; \quad j \ne k
\lab{pamjmzikb}\\
\sum_{j=1}^{m+1} \pa_{-j} m_{ik} &=& 0 \;\; ; \;\;
\sum_{j=1}^{m+1} \pa_{j} m_{ik} = 0 
\lab{summjBet}
\er
These relations couple to additional relations \rf{pajmzik},
\rf{pamjthmz} and \rf{pamjthmzik} involving the symmetric matrix 
$\theta^{(-1)}= ({\beta}_{ij})_{1\le i,j\le m+1}$ :
\br
\pa_{j}  m_{ik} &=&  {\beta}_{ij} m_{jk} 
\quad ; \quad i\ne j 
\lab{pajmzikg}\\
\pa_{j}  {\beta}_{ik}  &=&   {\beta}_{ij} 
{\beta}_{jk} 
\quad ; \quad i\, ,j \, ,k\, \;{\rm distinct}
\lab{pajthikg}\\
\pa_{-j}  {\beta}_{ik}  &=&    m_{ij}\, m_{kj} 
\lab{pamjthmzikg}
\er
with ${\beta}_{ij}$ satisfying :
\be
\sum_{j=1}^{m+1} \pa_{j} {\beta}_{ik} = 0 
\lab{upsik}
\ee
We notice a presence in our formalism of two Egoroff systems involving two
symmetric matrices ${\overline \cB}$ and $\theta^{(-1)}$.
The first one in \rf{pamjmzmoikb} and \rf{summjB} is realized in terms 
of the negative flows $u_{j}^{(-1)}$, while the second Egoroff system 
in \rf{pajthikg} and \rf{upsik} is based on the positive
flows $u_{j}^{(1)}$. Both systems are coupled to each other through the
matrix $M_0$.

The combined system of equations \rf{pamjmzmoikb}-\rf{upsik}
exhibits invariance under the simultaneous interchange :
\be
\pa_{-j} \leftrightarrow \pa_j \quad ; \quad {\bar \beta}_{ik} \leftrightarrow 
{\beta}_{ik}  \quad ; \quad m_{ij} \leftrightarrow  m_{ji}
\lab{syyme}
\ee
which maps one  Egoroff system into the other.
Equations \rf{pamjmzmoikb}-\rf{upsik} provide a coordinate form of the 
Cecotti-Vafa system \ct{Cecotti:1991me} which appeared in 
\ct{Dubrovin:1993yd,Razumov:1996me}.
The symmetry \rf{syyme} introduces a complex like structure 
analogous to complex conjugation on the complex manifold on which the
Cecotti-Vafa system was realized in \ct{Dubrovin:1993yd}.

For completeness we also list the identities :
\be 
\sum_{j=1}^{m+1} \pa_{j} (M_0^{-1} M_1) = \sum_{j=1}^{m+1} \pa_{j} {\overline \cB}
=I \;\; ;\;\;
\sum_{j=1}^{m+1} \pa_{j}  \theta^{(-1)} = I
\lab{sumjb}
\ee
which follow from \rf{paim0invm1} and \rf{pamjthmz}.

\subsection{Example: Reduction in Case of 
${\bf \widehat{gl} (2,\IC)}$-Hierarchy}

The ${ \widehat{gl} (2, \IC)}$-hierarchy contains the homogeneous ${ A_1}$-hierarchy 
(also known in the literature as the AKNS
hierarchy) together with a trivial decoupled scalar field.
Accordingly, we only consider the underlying ${ A_1}$-hierarchy.
In \ct{Aratyn:2000wr} the AKNS hierachy was extended
by the ``negative'' symmetry flows forming the 
Borel loop algebra.
 It was shown there how the complex sine-Gordon equation arises as a 
 symmetry flow
of the homogeneous ${ A_1}$-hierarchy. 
 The complex sine-Gordon and the Nonlinear Schr\"{o}dinger
equations appear as lowest negative and second
positive flows within the extended hierarchy.
Let $\cgh = \widehat{sl} (2)$ be a loop algebra on which we are given a graded
structure $ \cgh= \oplus_{n \in \IZ} \, \cgh_n$ with respect to an integral homogeneous
gradation defined by the  operator $d= \l d / d \l$.
The algebra $\cG = sl (2, \IC)$ has a standard basis 
$E_{\a} = \sigma_{+}$, $E_{-\a}= \sigma_{-}$ and $H = \sigma_{3}$.
We work within an algebraic approach to the integrable models based on
the linear spectral problem $L ( M )=0$ with a matrix Lax operator
containing the  matrix $A=q E_{\a} +r E_{-\a}$,\ct{Aratyn:1997ji}.
The second flow of the hierarchy:
\be
\pa_2 r = r_{xx}-2 q^2r \;\;\; ; \;\;\; \pa_2 q = -q_{xx}+2 q^2r 
\lab{secflow}
\ee
gives the familiar non-linear Schr\"{o}dinger equation.

The flow generated by $ E^{(-1)}$ is of special interest and we now
provide its zero-curvature formulation.
We choose the Gauss decomposition given by the following 
exponential of terms belonging to  zero grade subalgebra $\cgh_0 = sl (2)$ in 
order to parametrize $M_0$ satisfying (94):
\be
M_0 = e^{ \chi E_{-\a}}\, e^{R H} \,e^{ \psi E_{\a}}
\lab{B-def}
\ee
and define gauge potentials:
\be
\cA_{-} =  M_0 E^{(-1)} M_0^{-1} 
\;\; ;\;\; \cA_{+} =  - \pa_xM_0  M_0^{-1}+E^{(1)}
\lab{cap}
\ee
In order to match the number of independent modes in the matrix $A$ we impose
two ``diagonal'' constraints $\Tr \(\pa_x  M_0   M_0^{-1} H \) = 
\Tr \( M_0^{-1} { \pa}_{-1}  M_0  H \)=0$ which effectively eliminate
$R$ in terms of $\psi$ and $\chi$.
In fact, those constraints reduce the zero grade subspace $\cgh_0 = sl (2)$
into the coset $sl(2)/U(1)$.
In terms  of variables defined in \rf{B-def}, these constraints  determine 
the non-local field $R$ as
\br
 \pa_x R = { v \pa_x
u \o \Delta}, \quad \quad 
 { \pa}_{-1} R 
= { u { \pa}_{-1} v \o \Delta} \lab{bpar}
\er
where
\be
u = \psi \, e^R \quad ; \quad v = \chi \, e^R \quad ;
\quad \Delta = 1+ u\,v
\lab{tipc-def}
\ee
Since, $M_0$ has been chosen so that after imposition of the constraints
\rf{bpar} $\pa_x M_0 M_0^{-1} =  - \(q E_{\a} +r E_{-\a}\)$
we obtain the following representation for $q$ and $r$:
\be
q= - {(\pa_x u) \o \Delta} e^R \quad; \quad 
r = -(\pa_x v) \, e^{-R}
\lab{qr-dict}
\ee
and the zero-curvature condition \rf{toda-ov-id}: 
\be
\sbr{{ \pa}_{-1}+\cA_{-}}{\pa_x +\cA_{+}} =
{ \pa}_{-1} \cA_{+} - \pa_x \cA_{-} + \sbr{\cA_{-}}{\cA_{+}} =0
\lab{zcc}
\ee
leads to the equations of motion :
\br
{ \pa}_{-1} q &=& -{ \pa}_{-1} \( {\pa_x u \o \Delta} e^R \)
=   u e^R \lab{LRa}\\
{ \pa}_{-1} r &=&  - { \pa}_{-1} \(\pa_x v \, e^{-R}\)=
  v \Delta e^{-R} \lab{LRb}
\er

Let us now discuss the orthogonal reduction $-q= r=\P$  in expression
\rf{asigm}.  
This corresponds to setting $ v =-u$ and $e^{2R} = \Delta$  as follows
from equations \rf{bpar}-\rf{qr-dict}. Equations \rf{LRa} or \rf{LRb} become
in this limit :
\br
\pa_{-1} \pa_x u + {{ u\pa_x u \pa_{-1}u}\o {\Delta}} = -u \Delta
\lab{equ}
\er
Using the 2x2 representation of the $sl(2)$ algebra together with the constraint
$u = - v$ it follows from \rf{B-def}
that under the orthogonal reduction the matrix $M_0$  takes the following
form :
\br
M_0 =  \fourmat{e^{R}}{ u}{ -u }{ e^R },\quad 
{ M_0}^{-1} =  \fourmat{e^{R}}{ -u}{ u }{ e^R }
\lab{mzero}
\er
which reproduces formulas derived from the pseudo-differential appproach in 
equations \rf{mzeropseudo} and \rf{pppaa}-\rf{pppab}.

The constraint $M_0^{\dagger} = M_0$ amounts to choosing $u$ as a purely
imaginary function :
\be
u = i \sinh \b
\lab{sinhb}
\ee
In this parametrization $e^R= \cosh \b$.
Plugging this into equation \rf{equ} we obtain the sinh-Gordon equation :
\br
\pa_{-1}\pa_x \b = - \h  \sinh (2\b )
\lab{sg}
\er
for the reduced hierarchy.
One can verify that
$M_0^{-1} M_1$ becomes now a symmetric matrix
in agreement with equation \rf{paim0invm1}.

\vskip 10pt \noindent
{\bf Acknowledgements} \\
H.A. is partially supported by NSF (PHY-9820663).
The work of J.F.G and A.Z.
is supported in part by CNPq.
H.A. thanks Fapesp for support and IFT for hospitality during his visit.
We thank G. Sotkov for suggestion regarding the Cecotti-Vafa equations.

\end{document}